\documentclass[review]{elsarticle}

\usepackage{amssymb}
\usepackage{graphicx}
\usepackage{graphics}

\usepackage{lineno,hyperref}
\modulolinenumbers[5]

\journal{Applied Mathematics and Computation}

\begin{document}

\begin{frontmatter}

\title{The flow and heat transfer in a viscous fluid over an unsteady stretching surface}

\author[RDEne]{Remus-Daniel Ene\corref{mycorrespondingauthor}}
\cortext[mycorrespondingauthor]{Corresponding author.}
\ead{remus.ene@upt.ro}

\author[VMarinca]{Vasile Marinca}
\ead{vmarinca@mec.upt.ro}

\author[BMarinca]{Bogdan Marinca}
\ead{bogdan.marinca@upt.ro}

\address[RDEne]{University Politehnica Timi\c soara, Department of
Mathematics, Timi\c soara, 300006, Romania}
\address[VMarinca]{University Politehnica Timi\c soara, Department of Mechanics and Vibration, Timi\c soara, 300222, Romania\\
Department of Electromechanics and Vibration, Center for Advanced
and Fundamental Technical Research, Romania Academy, Timi\c soara,
300223, Romania}
\address[BMarinca]{University Politehnica Timi\c soara, Department of Applied Electronics, Timi\c soara, 300223, Romania}



%
%

\begin{abstract}
In this paper we have studied the flow and heat transfer in a
viscous fluid by a horizontal sheet. The stretching rate and
temperature of the sheet vary with time. The governing equations
for momentum and thermal energy are reduced to ordinary
differential equations by means of similarity transformation.
These equations are solved approximately by means of the Optimal
Homotopy Asymptotic Method (OHAM) which provides us with a
convenient way to control the convergence of approximation
solutions and adjust convergence rigorous when necessary. Some
examples are given and the results obtained reveal that the
proposed method is effective and easy to use.
\end{abstract}


\end{frontmatter}

\linenumbers

\section{Introduction}
\label{1}

\noindent \par The flow and heat transfer in a viscous fluid over
a stretching surface is a relevant problem in many industrial and
engineering processes. Examples are manufacture and drawing of
plastics and rubber sheets, polymer extrusion, wire drawing,
glass-fiber and paper production, crystal growing, continuous
casting, and so on. Cooling of stretching surface requires
dedicated control of the temperature and consequently knowledge of
flow and heat transfer in a viscous fluid. Sakiadis \cite{1},
\cite{2}, studied the boundary layer flow over a continuous solid
surface moving with constant speed. Crane \cite{3} analyzed the
stretching problem having in view the fluid flow over a linearly
stretching surface. Tsou et al \cite{4} studied constant surface
velocity and temperature. Gupta and Gupta \cite{5} and Maneschy et
al \cite{6} extended the Crane' work to the stretching problem
with a constant surface temperature including suction or blowing
and to fluids exhibiting a non-Newtonian behavior, respectively.
Grubka and Bobba \cite{7} studied the stretching problem for a
surface moving with a linear velocity and with a variable surface
temperature. Wang \cite{8} introduced a similarity transformation
to reduce time-dependent momentum equation to a third-order
nonlinear differential equation. He analyzed the hydrodynamic
behavior of a finite fluid body driven by an unsteady stretching
surface. The same problem was considered by Usha and Rukamani
\cite{9} for the axisymmetric case. Anderson et al \cite{10}
analyzed the accompanying heat transfer in the liquid film driven
by unsteady stretching surface. Ali \cite{11} and Magyari et al
\cite{12} considered permeable surfaces and different surface
temperature distributions. Vajravelu \cite{13} studied the flow
and heat transfer in a viscous fluid over a planar nonlinear
stretching sheet. Magyari and Keller \cite{14} applied the Merkin
transformation method to the heat transfer problems of steady
boundary layer flows induced by stretching surfaces. Elbashbeshy
and Bazid \cite{15} studied similarity solution of the laminar
boundary layer equations corresponding to an unsteady stretching
surface. Dandapat et al \cite{16} assumed that the stretching
surface is stretched impulsively from rest and the effect of
inertia of the liquid is considered. The unsteady heat and fluid
flow has been investigated by Ali and Magyari \cite{17}. Liu and
Anderson \cite{18} explored the thermal characteristics of a
viscous film on an unsteady stretching surface. Chen \cite{19}
analyzed the problem of MHD mixed convective flow and heat
transfer of an electrically conducting, power-low fluid past a
stretching surface in the presence of heat generation/absortion
and thermal radiation. Dandapat et al \cite{20} studied a thin
viscous liquid film flow over a stretching sheet under different
non-linear stretching velocities in presence of uniform transverse
magnetic field. Cortell \cite{21} presented momentum and heat
transfer for the flow induced in a quiscent fluid by a permeable
non-linear stretching sheet with a prescribed power-low
temperature distribution.

Analytical solutions to nonlinear differential equations play an
important role in the study of flow and heat transfer of different
types fluids, but it is difficult to find these solutions in the
presence of strong nonlinearity. A few approaches have been
proposed to find and develop approximate solutions of nonlinear
differential equations. Perturbation methods have been applied to
determine approximate solutions to weakly nonlinear problems
\cite{22}. But the use of perturbation theory in many problems is
invalid for parameters beyond a certain specified range. Other
procedures have been proposed such as the Adomian decomposition
method \cite{23}, some linearization methods \cite{24}, \cite{25}
, various modified Lindstedt-Poincare methods \cite{26},
variational iteration method \cite{27}, optimal homotopy
perturbation method \cite{28}, optimal homotopy asymptotic method
\cite{29} - \cite{33}.

In the present work we propose an accurate approach to nonlinear
differential equations of the flow and heat transfer in a viscous
fluid, using an analytical technique, namely optimal homotopy
asymptotic method. Our procedure, which does not imply the
presence of a small or large parameter in the equation or into the
boundary/initial conditions, is based on the construction and
determination of the linear operators and of the auxiliary
functions, combined with a convenient way to optimally control the
convergence of the solution. The efficiency of the proposed
procedure is proves while an accurate solution is explicitly
analytically obtained in an iterative way after only one
iteration. The validity of this method is demonstrated by
comparing the results obtained with the numerical solution.

\section{Equations of motion}
\label{sec:2}

Consider an unsteady, two dimensional flow on a continuous
stretching surface, with the governing time-dependent equations
for the continuity, momentum and thermal energy \cite{8},
\cite{10}, \cite{15}, \cite{17}, \cite{18}:
\begin{eqnarray}
\label{AliM1} \frac{\partial u}{\partial x} + \frac{\partial
v}{\partial y} = 0
\end{eqnarray}

\begin{eqnarray}
\label{AliM2} \frac{\partial u}{\partial t} + u \frac{\partial
u}{\partial x}+ v \frac{\partial v}{\partial y} = v
\frac{\partial^2 u}{\partial y^2}
\end{eqnarray}

\begin{eqnarray}
\label{AliM3} \frac{\partial T}{\partial t} + u \frac{\partial
T}{\partial x}+ v \frac{\partial T}{\partial y} = k
\frac{\partial^2 T}{\partial y^2}
\end{eqnarray}
where $u$ and $v$ are velocity components in the $x$ and $y$
directions, respectively, $T$ is the temperature and $k$ is the
thermal conductivity of the incompressible fluid. The appropriate
boundary conditions are:
\begin{eqnarray}\label{AliM4}
u = \frac{u_0 \frac{x}{l}}{1 + \gamma t}, \ \ \ v=0, \ \ \ T=
T_{\infty} + \frac{ T_0}{(1+ \gamma t)^c} \Big(\frac{x}{l}\Big)^n
\ \qquad \textrm{at} \ y=0
\end{eqnarray}

\begin{eqnarray}\label{AliM5}
u \rightarrow 0, \ \ \ T \rightarrow T_{\infty}  \ \qquad
\textrm{at} \ y\rightarrow \infty
\end{eqnarray}
where $u_0$, $T_0$, $T_{\infty}$, $\gamma$ are positive constants,
$c$ and $n$ are arbitrary and $l$ is a reference length.\\
If $Re=\frac{u_0 l}{v}$ and $Pr=\frac{v}{k}$ are the Reynolds
number and the Prandl number respectively and if we choose a
stream function $\Psi(x,y)$ such that
\begin{eqnarray}\label{AliM6}
u = \frac{\partial \Psi}{\partial y}, \ \quad \ v = -
\frac{\partial \Psi}{\partial x}
\end{eqnarray}
then the equation (\ref{AliM1}) of continuity is satisfied and the
mathematical analysis of the problems (\ref{AliM2}) and
(\ref{AliM3}) is simplified by introducing the following
similarity transformation:
\begin{eqnarray}\label{AliM7}
\Psi = \frac{x}{l} \frac{f(\eta)}{\sqrt{Re} (1+ \gamma t)^{1/2}}
\end{eqnarray}

\begin{eqnarray}\label{AliM8}
\eta = \sqrt{Re} \frac{y}{l (1+ \gamma t)^{1/2}}
\end{eqnarray}

\begin{eqnarray}\label{AliM9}
T =T_{\infty} + T_0 \Big(\frac{x}{l}\Big)^n
\frac{\theta(\eta)}{(1+ \gamma t)^{c}}
\end{eqnarray}
$T_0$ being a reference temperature. In this way Eqs.
(\ref{AliM6}) can be written in the form:
\begin{eqnarray}\label{AliM10}
u(x,y,t) = \frac{u_0}{l} \frac{x}{(1+\gamma t)} f'(\eta)
\end{eqnarray}
\begin{eqnarray}\label{AliM11}
v(x,y,t) =- \frac{u_0}{\sqrt{Re} (1 + \gamma t)^{1/2}} f(\eta)
\end{eqnarray}
where prime denotes differentiation with respect to $\eta$.

Substituting Eqs. (\ref{AliM7}), (\ref{AliM8}), (\ref{AliM9}),
(\ref{AliM10}) and (\ref{AliM11}) into Eqs. (\ref{AliM2}) and
(\ref{AliM3}), we obtain
\begin{eqnarray}\label{AliM12}
f'''+ff''-{f'}^2 + \Lambda \Big(f' + \frac{1}{2} \eta f''\Big)=0
\end{eqnarray}
\begin{eqnarray}\label{AliM13}
\frac{1}{Pr}\theta''+f\theta'-n{f'}\theta + \Lambda \Big(c \theta
+ \frac{1}{2} \eta \theta'\Big)=0.
\end{eqnarray}

Here $\Lambda = \frac{\gamma l}{u_0}$ is dimensionless measure of
the unsteadiness.

The dimensional boundary conditions (\ref{AliM4}) and
(\ref{AliM5}) become
\begin{eqnarray}\label{AliM14}
u= \frac{u_0}{l} \frac{x}{(1+\gamma t)} f'(0) \ \quad \
\textrm{at} \ y=0
\end{eqnarray}
\begin{eqnarray}\label{AliM15}
v =- \frac{u_0}{\sqrt{Re} (1 + \gamma t)^{1/2}} f(0) \ \quad \
\textrm{at} \ y=0
\end{eqnarray}
\begin{eqnarray}\label{AliM16}
T =T_{\infty} + T_0 \Big(\frac{x}{l}\Big)^n \frac{\theta(0)}{(1+
\gamma t)^{c}} \ \quad \ \textrm{at} \ y=0
\end{eqnarray}
such that for the dimensionless functions $f$ and $\theta$, the
boundary/initial conditions become
\begin{eqnarray}\label{AliM17}
f(0)=f_w, \ \ \ f'(0) = 1, \ \ \ f'(\infty)=0
\end{eqnarray}
\begin{eqnarray}\label{AliM18}
\theta(0)=1, \ \ \ \theta(\infty)=0.
\end{eqnarray}

In addition to the boundary conditions (\ref{AliM17}) and
(\ref{AliM18}), the requirements
\begin{eqnarray}\label{AliM19}
f'(\eta) \geq 0, \ \quad \ \theta(\eta) \geq 0, \ \  \forall \
\eta \geq 0
\end{eqnarray}
must also satisfied \cite{17}.

\section{Basic ideas of optimal homotopy asymptotic method}
\label{sec:3}

Eqs. (\ref{AliM12}) (or (\ref{AliM13})) with boundary conditions
(\ref{AliM17}) (or (\ref{AliM18})) can be written in a more
general form:
\begin{eqnarray}\label{AliM20}
N\Big(\Phi(\eta)\Big)=0
\end{eqnarray}
where $N$ is a given nonlinear differential operator depending on
the unknown function $\Phi(\eta)$, subject to the initial/boundary
conditions:
\begin{eqnarray}\label{AliM21}
B\Big(\Phi(\eta), \frac{d \Phi(\eta)}{d \eta}\Big)=0.
\end{eqnarray}

It is clear that $\Phi(\eta) =f(\eta)$ or $\Phi(\eta) =
\theta(\eta)$.

Let $\Phi_0(\eta)$ be an initial approximation of $\Phi(\eta)$ and
$L$ an  arbitrary linear operator such as
\begin{eqnarray}\label{AliM22}
L\Big(\Phi_0(\eta)\Big) =0, \ \ \ B\Big(\Phi_0(\eta), \frac{d
\Phi_0(\eta)}{d\eta}\Big)=0.
\end{eqnarray}

We remark that this operator $L$ is not unique.

If $p \in [0, \ 1]$ denotes an embedding parameter and $F$ is a
function, then we propose to construct a homotopy \cite{29} -
\cite{33}:
\begin{eqnarray}\label{AliM23}
\mathcal{H} \Big[ L\Big( F(\eta,p) \Big), \ H(\eta, C_i), \ N\Big(
F(\eta,p) \Big) \Big]
\end{eqnarray}
with the following two properties:
\begin{eqnarray}\label{AliM24}
\mathcal{H} \Big[ L\Big( F(\eta,0) \Big), \ H(\eta, C_i), \ N\Big(
F(\eta,0) \Big) \Big]= \nonumber \\
=L\Big( F(\eta,0) \Big) = L\Big( \Phi_0(\eta) \Big)
\end{eqnarray}
\begin{eqnarray}\label{AliM25}
\mathcal{H} \Big[ L\Big( F(\eta,1) \Big), \ H(\eta, C_i), \ N\Big(
F(\eta,1) \Big) \Big]= H(\eta, C_i) N\Big( \Phi(\eta) \Big)
\end{eqnarray}
where $H(\eta, C_i) \neq 0$, is an arbitrary auxiliary
convergence-control function depending on variable $\eta$ and on a
number of arbitrary parameters $C_1$, $C_2$, ..., $C_m$ which
ensure the convergence of the approximate solution.

Let us consider the function $F$ in the form
\begin{eqnarray}\label{AliM26}
F(x,p) = \Phi_0(\eta) + p \Phi_1(\eta, C_i) + p^2 \Phi_2(\eta,
C_i)+...
\end{eqnarray}

By substituting Eq. (\ref{AliM26}) into equation obtained by means
of the homotopy (\ref{AliM23})
\begin{eqnarray}\label{AliM27}
\mathcal{H} \Big[ L\Big( F(\eta,p) \Big), \ H(\eta, C_i), \ N\Big(
F(\eta,p) \Big) \Big] = 0
\end{eqnarray}
and equating the coefficients of like powers of $p$, we obtain the
governing equation of $\Phi_0(x)$ given by Eq. (\ref{AliM22}) and
the governing equation of $\Phi_1(\eta,C_i)$, $\Phi_2(\eta,C_i)$
and so on. If the series (\ref{AliM26}) is convergent at $p=1$,
one has:
\begin{eqnarray}\label{AliM28}
F(\eta,1) = \Phi_0(\eta) + \Phi_1(\eta, C_i) + \Phi_2(\eta,
C_i)+...
\end{eqnarray}

But in particular we consider only the first-order approximate
solution
\begin{eqnarray}\label{AliM29}
\overline{\Phi}(\eta, C_i) = \Phi_0(\eta) + \Phi_1(\eta, C_i), \ \
\ i=1,2,...,m
\end{eqnarray}
and the homotopy (\ref{AliM23}) in the form
\begin{eqnarray}\label{AliM30}
\mathcal{H} \Big[ L\Big( F(\eta,p) \Big), \ H(\eta, C_i), \ N\Big(
F(\eta,p) \Big) \Big] = L\Big(\Phi_0(\eta)\Big) + \nonumber \\
+ p \Big[ L\Big(\Phi_1(\eta, C_i)\Big) - L\Big(\Phi_0(\eta)\Big) +
H(\eta, C_i) N\Big( \Phi_0(\eta) \Big) \Big].
\end{eqnarray}

Equating only the coefficients of $p^0$ and $p^1$ into Eq.
(\ref{AliM30}), we obtain the governing equation of $\Phi_0(\eta)$
given by Eq. (\ref{AliM22}) and the governing equation of
$\Phi_1(\eta, C_i)$ i.e.
\begin{eqnarray}\label{AliM31}
L\Big(\Phi_1(\eta,C_i)\Big) = H(\eta, C_i) N\Big( \Phi_0(\eta)
\Big), \nonumber \\
B\Big(\Phi_1(\eta,C_i), \frac{d \Phi_1(\eta,C_i)}{d\eta}\Big)=0, \
\ i=1,2,...,m.
\end{eqnarray}

It should be emphasize that $\Phi_0(\eta)$ and $\Phi_1(\eta,C_i)$
are governed by the linear Eqs. (\ref{AliM22}) and (\ref{AliM31}),
respectively with boundary conditions that come from the original
problem, which can be easily solved. The convergence of the
approximate solution (\ref{AliM29})depends upon the auxiliary
convergence-control function $H(\eta, C_i)$. There are many
possibilities to choose the function $H(\eta, C_i)$. Basically,
the shape of $H(\eta, C_i)$ must follow the terms appearing in the
Eq. (\ref{AliM31}). Therefore, we try to choose $H(\eta, C_i)$ so
that in Eq. (\ref{AliM31}), the product $H(\eta, C_i) N\Big(
\Phi_0(\eta) \Big)$ be of the same shape with $N\Big( \Phi_0(\eta)
\Big)$. Now, substituting Eq. (\ref{AliM29}) into Eq.
(\ref{AliM20}), it results the following residual
\begin{eqnarray}\label{AliM32}
R(\eta,C_i) = N\Big( \overline{\Phi}(\eta,C_i) \Big).
\end{eqnarray}

At this moment, the first-order approximate solution given by Eq.
(\ref{AliM29}) depends on the parameters $C_1$, $C_2$, ..., $C_m$
and these parameters can be optimally identified via various
methods, such as the least square method, the Galerkin method, the
Kantorowich method, the collocation method or by minimizing the
square residual error:
\begin{eqnarray}\label{AliM33}
J(C_1,C_2,...,C_m) = \int_a^b R^2(\eta,C_1,C_2,...,C_m) \ d \eta
\end{eqnarray}
where $a$ and $b$ are two values depending on the given problem.
The unknown parameters $C_1$, $C_2$, ..., $C_m$ can be identified
from the conditions:
\begin{eqnarray}\label{AliM34}
\frac{\partial J}{\partial C_1} = \frac{\partial J}{\partial C_2}
= ... = \frac{\partial J}{\partial C_m} = 0.
\end{eqnarray}

With these parameters known (namely convergence-control
parameters), the first-order approximate solution (\ref{AliM29})
is well-determined.

\section{Application of OHAM to flow and heat transfer}
\label{sec:4}

We use the basic ideas of the OHAM by considering Eq.
(\ref{AliM12}) with the boundary conditions given by Eq.
(\ref{AliM17}). We can choose the linear operator in the form:\\
\begin{eqnarray}\label{AliM35}
L_f\Big({\Phi}(\eta)\Big) = \Phi''' - K^2 \Phi',
\end{eqnarray}
where $K>0$ is an unknown parameter at this moment.

We mention that the linear operator is not unique. Also, we have
freedom to choose:
$$L_{f}\Big(\Phi(\eta)\Big) = \Phi''' + \frac{3K}{K \eta + 1} \Phi''.$$

%
%
Eq. (\ref{AliM22}) becomes
\begin{eqnarray}
{\Phi_0}''' - K^2 {\Phi_0}' = 0, \ \ \Phi_0(0)=f_w, \ \
{\Phi_0}'(0)=1, \ \ {\Phi_0}'(\infty) = 0. \nonumber
\end{eqnarray}
which has the following solution
\begin{eqnarray}\label{AliM36}
{\Phi_0}(\eta) = f_w + \frac{1-e^{-K \eta}}{K}.
\end{eqnarray}

The nonlinear operator $N_f\Big({\Phi}(\eta)\Big)$ is obtained
from Eq. (\ref{AliM12}):
\begin{eqnarray}\label{AliM37}
N_f\Big({\Phi}(\eta)\Big) =
{\Phi}'''(\eta)+{\Phi}(\eta){\Phi}''(\eta)-{{\Phi}'(\eta)}^2 +
\Lambda \Big({\Phi}'(\eta) + \frac{1}{2} \eta
{\Phi}''(\eta)\Big)=0 \quad
\end{eqnarray}
such that substituting Eq. (\ref{AliM36}) into Eq. (\ref{AliM37}),
we obtain
\begin{eqnarray}\label{AliM38}
N_f\Big({\Phi_0}(\eta)\Big) =(\alpha \eta + \beta) e^{-K \eta}
\end{eqnarray}
where
\begin{eqnarray}\label{AliM39}
\alpha = \frac{1}{2} K \Lambda; \ \quad \ \beta =K^2-1-K f_w -
\Lambda.
\end{eqnarray}

Heaving in view that in Eq. (\ref{AliM38}) appears an exponential
function and that the auxiliary function $H_f(\eta, C_i)$ must
follow the terms appearing in Eq. (\ref{AliM38}), then we can
choose the function $H_f(\eta, C_i)$ in the following forms:
\begin{eqnarray}\label{AliM40}
H_f(\eta, C_i) = C_1 +C_2 \eta + (C_3 + C_4 \eta)e^{-K\eta} + (C_5
+ C_6 \eta)e^{-2 K \eta}
\end{eqnarray}
or
\begin{eqnarray}\label{AliM41}
H_f^{*}(\eta, C_i) = C_1  + (C_2 + C_3 \eta + C_4
\eta^2)e^{-K\eta}
\end{eqnarray}
or yet
\begin{eqnarray}\label{AliM42}
H_f^{**}(\eta, C_i) = C_1 + C_2 \eta + C_3 \eta^2  + (C_4 + C_5
\eta )e^{-K\eta} + \nonumber \\
+(C_6 + C_7 \eta + C_8 \eta^2)e^{-2 K \eta}
\end{eqnarray}
and so on, where $C_1$, $C_2$, ... are unknown parameters at this
moment.

If we choose only the expression (\ref{AliM40}) for $H_f(\eta,
C_i)$, then by using Eqs. (\ref{AliM38}),
 (\ref{AliM40}) and (\ref{AliM31}), we can obtain the equation in $\Phi_1(\eta,
 C_i)$:
\begin{eqnarray}\label{AliM43}
{\Phi_1}''' - K^2 {\Phi_1}' = \Big[\beta C_1 + (\alpha C_1 + \beta
C_2)\eta + \alpha C_2 \eta^2 \Big] e^{-K \eta} + \nonumber \\
+ \Big[\beta C_3 + (\alpha C_3 + \beta C_4)\eta + \alpha C_4
\eta^2 \Big] e^{-2 K \eta} + \Big[\beta C_5 + (\alpha C_5 + \beta
C_6)\eta + \nonumber \\
+ \alpha C_6 \eta^2 \Big] e^{-3 K \eta}, \ \ \Phi_1(0)=
{\Phi_1}'(0)={\Phi_1}'(\infty) = 0. \quad \quad
\end{eqnarray}

The solution of Eq. (\ref{AliM43}) can be found as
\begin{eqnarray}\label{AliM44}
{\Phi_1}(\eta) = M_1+ \Big[N_1 + \Big( \frac{7 \alpha C_2 }{4 K^4}
+ \frac{3 \alpha C_1 }{4 K^3} + \frac{3 \beta C_2 }{4 K^3} +
\frac{\beta C_1 }{2 K^2} \Big)\eta + \nonumber \\
+ \Big( \frac{3 \alpha C_2 }{4 K^3} + \frac{\alpha C_1 }{4 K^2} +
 \frac{\beta C_2 }{4 K^2} \Big)\eta^2 + \frac{\alpha C_2 }{6 K^2} \eta^3
 \Big] e^{-K \eta} + \Big[- \frac{85 \alpha C_4 }{108 K^5}
- \frac{11 \alpha C_3 }{36 K^4} - \nonumber \\
- \frac{11 \beta C_4 }{36 K^4} - \frac{\beta C_3 }{6 K^3} - \Big(
\frac{11 \alpha C_4 }{18 K^4} + \frac{\alpha C_3 }{6 K^3} +
\frac{\beta C_4 }{6 K^3} \Big)\eta -\frac{\alpha C_4 }{6 K^3} \eta^2 \Big] e^{-2 K \eta} + \nonumber \\
 + \Big( -
\frac{115 \alpha C_6 }{1728 K^5} - \frac{13 \alpha C_5 }{288 K^4}
- \frac{13 \beta C_6 }{288 K^4} - \frac{\beta C_5 }{24 K^3} \Big)
e^{-K \eta} \quad \quad
\end{eqnarray}
where
\begin{eqnarray}\label{AliM45}
 M_1 = -\frac{3 \alpha + 2K \beta}{4 K^4} C_1 -\frac{7 \alpha + 3 K \beta}{4 K^5} C_2-\frac{5 \alpha + 6K \beta}{36 K^4}
 C_3- \nonumber\\
 -\frac{19 \alpha + 15 K \beta}{108 K^5} C_4-\frac{7 \alpha + 12 K \beta}{144 K^4} C_5-\frac{37 \alpha + 42 K \beta}{864 K^5}
 C_6 \nonumber \\
 N_1 =  \frac{3 \alpha + 2 K \beta}{4 K^4} C_1 + \frac{7 \alpha + 3 K \beta}{4 K^5} C_2 + \frac{4 \alpha + 3 K \beta}{9 K^4}
 C_3+ \nonumber\\
 +\frac{26 \alpha + 12 K \beta}{27 K^5} C_4 + \frac{3 \alpha + 4 K \beta}{32 K^4} C_5 + \frac{7 \alpha + 6 K \beta}{64 K^5}
 C_6.
 \end{eqnarray}

 The first-order approximate solution (\ref{AliM29}) for Eqs. (\ref{AliM12}) and (\ref{AliM17}) is obtained
 from Eqs. (\ref{AliM36}) and (\ref{AliM45}):
\begin{eqnarray}\label{AliM46}
 \overline{f}(\eta) = \overline{\Phi}(\eta) = {\Phi_0}(\eta) +
 {\Phi_1}(\eta).
 \end{eqnarray}

 In what follows, we consider Eqs. (\ref{AliM13}) and
 (\ref{AliM18}). In this case, we choose the linear operator in
 the form
\begin{eqnarray}\label{AliM47}
L_{\theta}\Big(\varphi(\eta)\Big) = \varphi''+K \varphi'
 \end{eqnarray}
 where the parameter $K$ is defined in Eq. (\ref{AliM35}).

 Eq. (\ref{AliM22}) becomes
\begin{eqnarray}\label{AliM48}
{\varphi_0}''+K {\varphi_0}'=0, \ \ \ \varphi_0(0)=1, \ \ \
\varphi_0(\infty)=0.
 \end{eqnarray}

Eq. (\ref{AliM48}) has the solution
\begin{eqnarray}\label{AliM49}
\varphi_0(\eta)=e^{-K \eta}.
 \end{eqnarray}

 The nonlinear operator $N_{\theta}\Big(\varphi(\eta)\Big) $ is
 obtained from Eq. (\ref{AliM12}):
\begin{eqnarray}\label{AliM50}
N_{\theta}\Big(\varphi(\eta)\Big) = \frac{1}{Pr}\varphi''+ \Phi
\varphi'-n{\Phi'}\varphi + \Lambda \Big(c \varphi + \frac{1}{2}
\eta \varphi' \Big).
\end{eqnarray}

Substituting Eq. (\ref{AliM49}) into Eq. (\ref{AliM50}), we obtain
\begin{eqnarray}\label{AliM51}
N_{\theta}\Big(\varphi_0(\eta)\Big) = (m_1 \eta + m_2) e^{- K
\eta} + m_3 e^{-2 K \eta}
\end{eqnarray}
where
\begin{eqnarray}\label{AliM52}
m_1 =- \frac{1}{2} K \Lambda; \ \ \ m_2 = \frac{K^2}{Pr} - K f_w
-1 + c \Lambda; \ \ \ m_3 = 1-n.
\end{eqnarray}

The auxiliary function $H_{\theta}(\eta, C_i)$ can be choose in
the forms:
\begin{eqnarray}\label{AliM53}
H_{\theta}(\eta, C_i) = C_7 +C_8 \eta + (C_9 + C_{10}
\eta)e^{-K\eta} + (C_{11} + C_{12} \eta)e^{-2 K \eta}
\end{eqnarray}
or
\begin{eqnarray}\label{AliM54}
H_{\theta}^{*}(\eta, C_i) = C_7 + C_8 \eta + C_9 \eta^2  + (C_{10}
+ C_{11} \eta )e^{-K\eta} +  C_{13} e^{-2 K \eta}
\end{eqnarray}
or yet
\begin{eqnarray}\label{AliM55}
H_{\theta}^{**}(\eta, C_i) =C_7  + (C_8 + C_9 \eta)e^{-K\eta}  +
(C_{10} + C_{11} \eta)e^{-2 K \eta}
\end{eqnarray}
and so on, where $C_7$, $C_8$, ... are unknown parameters.

If we choose the Eq. (\ref{AliM53}) for $H_{\theta}$, then from
Eqs. (\ref{AliM51}), (\ref{AliM53}) and (\ref{AliM31}) we obtain
the equation in $\varphi_1(\eta, C_i)$ as
\begin{eqnarray}\label{AliM56}
{\varphi_1}''+K {\varphi_1}' = \Big[ m_2 C_7 + (m_1 C_7 + m_2 C_8)
\eta
+ m_1 C_8 \eta^2 \Big] e^{- K \eta} + \nonumber \\
+ \Big[ m_2 C_9 + m_3 C_7 + (m_1 C_9 + m_2 C_{10} + m_3 C_8) \eta
+ m_1 C_{10} \eta^2 \Big] e^{- 2 K \eta} + \nonumber \\
+ \Big[ m_3 C_9 + m_2 C_{11} + (m_3 C_{10} + m_1 C_{11} + m_2
C_{12}) \eta + m_1 C_{12} \eta^2 \Big] e^{- 3 K \eta} + \nonumber
\\
+ (m_3 C_{11} + m_3 C_{12} \eta) e^{-4 K \eta}, \ \ \ \phi_1(0)=
\phi_1(\infty)=0. \quad \quad
\end{eqnarray}

Solving Eq. (\ref{AliM56}), we obtain
\begin{eqnarray}\label{AliM57}
\varphi_1(\eta)  =  \Big[ P_1 - \Big( \frac{2 m_{1} C_8}{K^3} +
\frac{m_{1} C_7}{K^2}+ \frac{m_{2} C_8}{K^2} + \frac{m_{2} C_7}{K}
\Big) \eta - \Big( \frac{m_{1} C_8}{K^2} + \frac{m_{1} C_7}{2 K}+
\nonumber \\
+ \frac{m_{2} C_8}{2 K } \Big) \eta^2 - \frac{m_{1} C_8}{3 K }
\eta^3 \Big] e^{- K \eta} +
 \Big[ \frac{7 m_1 C_{10}}{4 K^4} + \frac{3 m_1 C_{9}}{4 K^3} + \frac{3 m_2 C_{10}}{4 K^3} + \frac{3 m_3 C_{8}}{4 K^3}
 + \nonumber \\
+ \frac{ m_2 C_{9}}{2 K^2} + \frac{ m_3 C_{7}}{2 K^2} + \Big(
\frac{3 m_{1} C_{10}}{2 K^3} + \frac{m_{1} C_9}{2 K^2}+
\frac{m_{2} C_{10}}{2 K^2} + \frac{m_{3} C_8}{2K^2} \Big) \eta + \nonumber \\
+ \frac{m_{1} C_{10}}{2 K^2 } \eta^2 \Big] e^{- 2 K \eta} + \Big[
\frac{5(m_3 C_{10} + m_1 C_{11} + m_2 C_{12})}{36 K^3}  + \frac{
m_3 C_{9} + m_2 C_{11}}{6 K^2} + \nonumber \\
+\Big(  \frac{m_3 C_{10} + m_1 C_{11} + m_2 C_{12}}{6 K^2} +
\frac{5 m_{1} C_{12}}{18 K^3} \Big) \eta + \frac{m_{1} C_{12}}{6
K^2} \eta^2 \Big] e^{- 3 K \eta} + \nonumber \\
+ \Big( \frac{7 m_{3} C_{11}}{12 K^2} + \frac{7 m_{3} C_{12}}{144
K^3} + \frac{ m_{3} C_{12}}{12 K^2} \eta) e^{-4 K \eta}, \ \ \
\phi_1(0)= \phi_1(\infty)=0, \quad \quad \quad
\end{eqnarray}
where
\begin{eqnarray}\label{AliM58}
P_1=- \frac{m_3 C_{7}}{2 K^2} - \frac{3 m_3 C_{8}}{4 K^3} -
\frac{9 m_1 + 6 K m_2 + 2 K m_3}{12 K^3} C_9 -\nonumber \\
- \frac{63 m_1 + 27 K m_2 + 5 K m_3}{36 K^4} C_{10} - \frac{5 m_1
+  K (m_2 + 3 m_3)}{36 K^3} C_{11} - \nonumber \\
- \frac{20 m_2 + 7 m_3}{144 K^3} C_{12}.
\end{eqnarray}

In this way, the first-order approximate solution (\ref{AliM29})
for Eqs. (\ref{AliM13}) and (\ref{AliM18}) becomes
\begin{eqnarray}\label{AliM59}
\overline{\theta}(\eta) = \overline{\varphi}(\eta) =
\varphi_0(\eta) + \varphi_1(\eta,C_i).
\end{eqnarray}

\section{Numerical examples}
\label{sec:5}

In order to prove the accuracy of the obtained results, we will
determine the convergence-control parameters $K$ and $C_i$ which
appear in Eqs. (\ref{AliM46}), (\ref{AliM59}) 
by means of the least square method. In this way, the
convergence-control parameters are optimally determined and the
first-order approximate solutions known for different values of
the known parameters $f_w$, $\Lambda$, $Pr$, $n$ and $c$. In what
follows, we illustrate the accuracy of the OHAM comparing
previously obtained approximate solutions  with the numerical
integration results computed by means of the shooting method
combined with fourth-order Runge-Kutta method using Wolfram
Mathematica 6.0 software. For some values of the parameters $f_w$,
$\Lambda$, $Pr$, $n$ and $c$ we will determine the approximate
solutions.

{\textbf{Example 5.1.a}} For the first alternative given in the
subsection 4.1, we consider $f_w=-1$, $\Lambda=1$,
$c=\frac{1}{2}$, $n=1$, $Pr=0.7$. For Eq. (\ref{AliM46}),
following the procedure described above are obtained the
convergence-control parameters:
\begin{eqnarray}
C_1 = -0.0881661632, \ C_2 = 0.0159074525, \
C_3 = 101.5499315816, \nonumber \\
C_4 = -16.3157319695, \ C_5 = -99.5951678657, \
C_6 = -64.5910051875 \nonumber \\
    K = 0.7591636981 \nonumber
\end{eqnarray}
and consequently the first-order approximate solution
(\ref{AliM46}) can be written in the form:
\begin{eqnarray}\label{AliM77}
\overline{f}(\eta) = 0.4921333156  +  (-0.5668554881
   + 0.0071536463 \eta - \nonumber \\
-    0.0087518103 \eta^2 + 0.0017461596  \eta^3)
 e^{-0.7591636981 \eta} +  (-0.3980837570  - \nonumber \\
   - 7.4190413787 \eta +
     2.3591440003 \eta^2) e^{-1.5183273963 \eta} + (-0.5271940704  + \nonumber \\
   + 6.1764503488 \eta +
     2.3348551362 \eta^2) e^{-2.2774910945  \eta} \qquad
\end{eqnarray}


Now, for Eq. (\ref{AliM59}), the convergence-control parameters
are:
\begin{eqnarray}
     C_7 = 0.0363993085, \ C_8 = 0.0363993085, \
  C_9 =   -7.1448075448, \nonumber \\
 C_{10} =  4.3237724702, \ C_{11} = 42.1871800319, \
C_{12} =  18.0805214975 \nonumber
\end{eqnarray}
and therefore  the first-order approximate solution (\ref{AliM59})
becomes:
\begin{eqnarray}\label{AliM78}
\overline{\theta}(\eta) =   (0.3541683003 +
  0.2415876957 \eta - 0.0823906873 \eta^2 + \nonumber \\
+ 0.0060665514 \eta^3) e^{-0.7591636981
    \eta} +  (-2.6848440528 + \nonumber \\
   + 6.3660521866 \eta
-    1.4238603876 \eta^2) e^{-1.5183273963 \eta} + \nonumber \\
  +(3.3306757524 -
    3.3281240073 \eta -
    1.9846972772 \eta^2) e^{-2.2774910945 \eta} \quad \quad
\end{eqnarray}

In Tables 1 and 2 we present a comparison between the first-order
approximate solutions given by Eqs. (\ref{AliM77}) and
(\ref{AliM78}) respectively, with numerical results for some
values of variable $\eta$ and the corresponding relative errors.


\begin{table}[h!]
\caption{Comparison between OHAM results given by Eq.
(\ref{AliM77}) and numerical results for $f_w = -1$,
$\Lambda = 1$ }%
\label{tab:1}
\begin{tabular}{lllll}
 \hline\noalign{\smallskip}
  $\eta$ & $f_{\textrm{numeric}}$ & $ \overline{f}_{\textrm{OHAM}}$, Eq. (\ref{AliM77})& $\begin{array}{c}
                                                         \textrm{relative error}= \\
                                                          |f_{\textrm{numeric}}-\overline{f}_{\textrm{OHAM}}|
                                                          \end{array}$\\
   \noalign{\smallskip}\hline\noalign{\smallskip}
0   &  -1              &  -0.9999999999   &   1.88 $\cdot 10^{-15}$ \\   %
1   &  -0.1497942276   &  -0.1501421749   &    3.47 $\cdot 10^{-4}$ \\   %
2   &   0.3108643384   &   0.3106655367   &    1.98 $\cdot 10^{-4}$ \\   %
3   &   0.4604991620   &   0.4600705386   &    4.28 $\cdot 10^{-4}$ \\   %
4   &   0.4887865463   &   0.4894195278   &    6.32 $\cdot 10^{-4}$ \\   %
5   &   0.4919308455   &   0.4920394273   &    1.08 $\cdot 10^{-4}$ \\   %
6   &   0.4921388939   &   0.4918417250   &    2.97 $\cdot 10^{-4}$ \\   %
7   &   0.4921471111   &   0.4919782168   &    1.68 $\cdot 10^{-4}$ \\   %
8   &   0.4921472622   &   0.4922151064   &    6.78 $\cdot 10^{-5}$ \\   %
9   &   0.4921472290   &   0.4923440748   &    1.96 $\cdot 10^{-4}$ \\   %
10  &   0.4921472001   &   0.4923640175   &    2.16 $\cdot 10^{-4}$ \\   %
\noalign{\smallskip}\hline
    \end{tabular}
\end{table}
\begin{table}[h!]
\caption{Comparison between OHAM results given by Eq.
(\ref{AliM78}) and numerical results for $f_w = -1$,
$\Lambda = 1$, $c = \frac{1}{2}$, $n= 1$, $Pr=0.7$ }%
\label{tab:2}
\begin{tabular}{lllll}
 \hline\noalign{\smallskip}
  $\eta$ & ${\theta}_{\textrm{numeric}}$ & $ \overline{\theta}_{\textrm{OHAM}}$, Eq. (\ref{AliM78})& $\begin{array}{c}
                                                         \textrm{relative error}= \\
                                                          |{\theta}_{\textrm{numeric}}-\overline{\theta}_{\textrm{OHAM}}|
                                                          \end{array}$\\
   \noalign{\smallskip}\hline\noalign{\smallskip}
0   &    1             &    0.9999999999   &   8.88 $\cdot 10^{-16}$ \\   %
1   &   0.5325816311   &    0.5344078544   &    1.82 $\cdot 10^{-3}$ \\   %
2   &   0.2137609331   &    0.2123010871   &    1.45 $\cdot 10^{-3}$ \\   %
3   &   0.0624485224   &    0.0627998626   &    3.51 $\cdot 10^{-4}$ \\   %
4   &   0.0129724736   &    0.0141235139   &    1.15 $\cdot 10^{-3}$ \\   %
5   &   0.0019027817   &    0.0018865990   &    1.61 $\cdot 10^{-5}$ \\   %
6   &   0.0001968219   &   -0.0002871987   &    4.84 $\cdot 10^{-4}$ \\   %
7   &   0.0000144329   &   -0.0002516924   &    2.66 $\cdot 10^{-4}$ \\   %
8   &   8.27 $\cdot 10^{-7}$   &   0.0000468974   &    4.60 $\cdot 10^{-5}$ \\   %
9   &   1.08 $\cdot 10^{-7}$   &   0.0002281868   &    2.28 $\cdot 10^{-4}$ \\   %
10  &   7.47 $\cdot 10^{-8}$   &   0.0002807824   &    2.80 $\cdot 10^{-4}$ \\   %
\noalign{\smallskip}\hline
    \end{tabular}
\end{table}

{\textbf{Example 5.1.b}} In this case, we consider $f_w=-1$,
$\Lambda=1$, $c=\frac{1}{2}$, $n=1$, $Pr=2$. The solution
$\overline{f}(\eta)$ is given by Eq. (\ref{AliM77}). The
convergence-control parameters for Eq. (\ref{AliM59}) are:
\begin{eqnarray}
     C_7 = 0.3992391297, \ C_8 =  -0.0398233823, \         C_9 =   13.9195482545,  \nonumber \\
C_{10} =  -9.8140323466, \  C_{11} =  -37.1866029355, \ C_{12} =  -48.0634410813  \nonumber 
\end{eqnarray}
such that the first-order approximate solution (\ref{AliM59})
becomes:
\begin{eqnarray}\label{AliM79}
\overline{\theta}(\eta) =  (-2.1207041376 -           0.0561685488 \eta   + 0.0879369070 \eta^2 -  \nonumber \\
-   0.0066372303 \eta^3) e^{-0.7591636981 \eta} + e^{-1.5183273963   \eta} (7.9716346529 -  \nonumber \\
-   9.1921128520 \eta    + 3.2318564396 \eta^2)
e^{-0.7591636981 \eta} + (-4.8509305153  +  \nonumber \\
 +  8.0572360536 \eta +    5.2759197606 \eta^2) e^{-2.2774910945 \eta}  \quad \quad
\end{eqnarray}

In Table 3 we present a comparison between the first-order
approximate solutions given by Eq. (\ref{AliM79}) with numerical
results and corresponding relative errors.

\begin{table}[h!]
\caption{Comparison between OHAM results given by Eq.
(\ref{AliM79}) and numerical results for $f_w = -1$,
$\Lambda = 1$, $c = \frac{1}{2}$, $n= 1$, $Pr=2$ }%
\label{tab:3}
\begin{tabular}{lllll}
 \hline\noalign{\smallskip}
  $\eta$ & ${\theta}_{\textrm{numeric}}$ & $ \overline{\theta}_{\textrm{OHAM}}$, Eq. (\ref{AliM79})& $\begin{array}{c}
                                                         \textrm{relative error}= \\
                                                          |{\theta}_{\textrm{numeric}}-\overline{\theta}_{\textrm{OHAM}}|
                                                          \end{array}$\\
   \noalign{\smallskip}\hline\noalign{\smallskip}
0   &   1              &    0.9999999999   &   1.77 $\cdot 10^{-15}$ \\   
1   &   0.3341908857   &    0.3295767993   &    4.61 $\cdot 10^{-3}$ \\   
2   &   0.0347540513   &    0.0372485103   &    2.49 $\cdot 10^{-3}$ \\   
3   &   0.0011305745   &   -0.0002320988   &    1.36 $\cdot 10^{-3}$ \\   
4   &   0.0000127262   &   -0.0002852788   &    2.98 $\cdot 10^{-4}$ \\   
5   &   -5.44 $\cdot 10^{-8}$   &    0.0002991504   &    2.99 $\cdot 10^{-4}$ \\   
6   &   -9.15 $\cdot 10^{-8}$   &    0.0002884740   &    2.88 $\cdot 10^{-4}$ \\   
7   &   -7.96 $\cdot 10^{-8}$   &    0.0001372878   &    1.37 $\cdot 10^{-4}$ \\   
8   &   -7.05 $\cdot 10^{-8}$   &   -0.0000295212   &    2.94 $\cdot 10^{-5}$ \\   
9   &   -6.53 $\cdot 10^{-8}$   &   -0.0001505702   &    1.50 $\cdot 10^{-4}$ \\   
10  &   -5.99 $\cdot 10^{-8}$   &   -0.0002044072   &    2.04 $\cdot 10^{-4}$ \\   
\noalign{\smallskip}\hline
    \end{tabular}
\end{table}


{\textbf{Example 5.2.a}} For $f_w=0$, $\Lambda=1$,
$c=\frac{1}{2}$, $n=1$, $Pr=0.7$, the convergence-control
parameters for Eq. (\ref{AliM46}) are:
\begin{eqnarray}
   C_1 = -1.2640611927, \ C_2 =  0.1680009020, \
 C_3 =  -34.0575215187, \nonumber \\
  C_4 =  30.7898356526, \
  C_5 =  37.1281425060, \ C_6 =  13.8590545976,  \nonumber \\
     K =  1.1203766872 \nonumber
\end{eqnarray}
and therefore, the first-order approximate solution (\ref{AliM46})
can be written in the form:
\begin{eqnarray}\label{AliM80}
\overline{f}(\eta) = 0.9662722752   +  (1.3563995648 +
0.0351604322 \eta -  \nonumber \\
-    0.1157605059 \eta^2 +
     0.0124958644 \eta^3)  e^{-1.1203766872
\eta}  + (-2.5490820161 -   \nonumber \\
  - 1.7111113870 \eta -
    2.0440805123 \eta^2) e^{-2.2407533744 \eta} + (0.2264101761 -  \nonumber \\
  - 0.7552406743 \eta -
    0.2300192808 \eta^2) e^{-3.3611300616
\eta} \quad \quad
\end{eqnarray}

For Eq. (\ref{AliM59}), the convergence-control parameters are:
\begin{eqnarray}
  C_7= -1.6947892627, \
 C_8 =  0.2632100295, \
 C_9 = -1.2815579214,
 \nonumber \\
 C_{10} =  2.6268699384, \ C_{11} =  15.8897673738, \
 C_{12} =  9.5966071873 \nonumber
\end{eqnarray}
and the first-order approximate solution (\ref{AliM59}) is:
\begin{eqnarray}\label{AliM81}
\overline{\theta}(\eta) = (0.2987405005 + 1.1383970916 \eta -
        0.4581387031 \eta^2 +  \nonumber \\
    +   0.0438683382 \eta^3) e^{-1.1203766872  \eta} +  (-0.1000278783  +  \nonumber \\
      + 1.4818560845 \eta -
        0.5861577557 \eta^2) e^{-2.2407533744
\eta}  +  (0.8012873778 -  \nonumber \\
-    0.5959066165 \eta -
     0.7137932043 \eta^2) e^{-3.3611300616 \eta} \quad \quad
\end{eqnarray}

In Tables 4 and 5 we present a comparison between the first-order
approximate solutions given by Eqs. (\ref{AliM80}) and
(\ref{AliM81}) respectively, with numerical results and
corresponding relative errors.

\begin{table}[h!]
\caption{Comparison between OHAM results given by Eq.
(\ref{AliM80}) and numerical results for $f_w =0$,
$\Lambda = 1$ }%
\label{tab:4}
\begin{tabular}{lllll}
 \hline\noalign{\smallskip}
  $\eta$ & ${f}_{\textrm{numeric}}$ & $ \overline{f}_{\textrm{OHAM}}$, Eq. (\ref{AliM80})& $\begin{array}{c}
                                                         \textrm{relative error}= \\
                                                          |{f}_{\textrm{numeric}}-\overline{f}_{\textrm{OHAM}}|
                                                          \end{array}$\\
   \noalign{\smallskip}\hline\noalign{\smallskip}
0   &   -5.50 $\cdot 10^{-21}$   &   4.44 $\cdot 10^{-16}$   &    4.44 $\cdot 10^{-16}$ \\   
1   &   0.6894348341   &   0.6894914970   &   5.66 $\cdot 10^{-5}$ \\   
2   &   0.9167696529   &   0.9166682157   &   1.01 $\cdot 10^{-4}$ \\   
3   &   0.9608821303   &   0.9609858144   &   1.03 $\cdot 10^{-4}$ \\   
4   &   0.9659196704   &   0.9659030730   &   1.65 $\cdot 10^{-5}$ \\   
5   &   0.9662619960   &   0.9661631962   &   9.87 $\cdot 10^{-5}$ \\   
6   &   0.9662759513   &   0.9662663279   &   9.62 $\cdot 10^{-6}$ \\   
7   &   0.9662762950   &   0.9663395358   &   6.32 $\cdot 10^{-5}$ \\   
8   &   0.9662763018   &   0.9663501433   &   7.38 $\cdot 10^{-5}$ \\   
9   &   0.9662763032   &   0.9663306688   &   5.43 $\cdot 10^{-5}$ \\   
10  &   0.9662763043   &   0.9663080318   &   3.17 $\cdot 10^{-5}$ \\   
\noalign{\smallskip}\hline
    \end{tabular}
\end{table}

\begin{table}[h!]
\caption{Comparison between OHAM results given by Eq.
(\ref{AliM81}) and numerical results for $f_w = 0$,
$\Lambda = 1$, $c = \frac{1}{2}$, $n= 1$, $Pr=0.7$ }%
\label{tab:5}
\begin{tabular}{lllll}
 \hline\noalign{\smallskip}
  $\eta$ & ${\theta}_{\textrm{numeric}}$ & $ \overline{\theta}_{\textrm{OHAM}}$, Eq. (\ref{AliM81})& $\begin{array}{c}
                                                         \textrm{relative error}= \\
                                                          |{\theta}_{\textrm{numeric}}-\overline{\theta}_{\textrm{OHAM}}|
                                                          \end{array}$\\
   \noalign{\smallskip}\hline\noalign{\smallskip}
0   &   1              &   1.00           &   2.22 $\cdot 10^{-16}$ \\   
1   &   0.4003127445   &   0.4006174190   &   3.04 $\cdot 10^{-4}$ \\   
2   &   0.1184290061   &   0.1183366534   &   9.23 $\cdot 10^{-5}$ \\   
3   &   0.0250358122   &   0.0254649591   &   4.29 $\cdot 10^{-4}$ \\   
4   &   0.0037386526   &   0.0032572108   &   4.81 $\cdot 10^{-4}$ \\   
5   &   0.0003935287   &  -0.0000242886   &   4.17 $\cdot 10^{-4}$ \\   
6   &   0.0000291963   &   0.0001165652   &   8.73 $\cdot 10^{-5}$ \\   
7   &   1.53 $\cdot 10^{-6}$   &   0.0003370019   &   3.35 $\cdot 10^{-4}$ \\   
8   &   6.39 $\cdot 10^{-8}$   &   0.0003255701   &   3.25 $\cdot 10^{-4}$ \\   
9   &   8.43 $\cdot 10^{-9}$   &   0.0002261157   &   2.26 $\cdot 10^{-4}$ \\   
10  &   6.39 $\cdot 10^{-9}$   &   0.0001326393   &   1.32 $\cdot 10^{-4}$ \\    %
\noalign{\smallskip}\hline
    \end{tabular}
\end{table}


{\textbf{Example 5.2.b}} For $f_w=0$, $\Lambda=1$,
$c=\frac{1}{2}$, $n=1$, $Pr=2$ the firs-order approximate solution
(\ref{AliM46}) is given by Eq. (\ref{AliM80}). The
convergence-control parameters for Eq. (\ref{AliM59}) are
determined as:
\begin{eqnarray}
  C_7 = 0.4652101281, \
 C_8 = -0.0724620728, \ C_9 = 14.4924736065, \nonumber
 \\
 C_{10} =  -12.1643720147, \ C_{11} = 0.7277740132, \
 C_{12} =  -56.0450834796  \nonumber
\end{eqnarray}
such that the first-order approximate solution (\ref{AliM59}) may
be written as:
\begin{eqnarray}\label{AliM82}
\overline{\theta}(\eta) = (-1.4030799687  +
      0.1042610535 \eta +
      0.0880913412 \eta^2 -  \nonumber \\
    - 0.0120770121 \eta^3) e^{-1.1203766872  \eta} +
     (3.1476673319 -  \nonumber \\
-     3.8089261057 \eta +
      2.7143486992 \eta^2) e^{-2.2407533744 \eta} +  (-0.7445873632  +  \nonumber \\
        +  5.1973912018 \eta +
           4.1686190694 \eta^2) e^{-3.3611300616 \eta} \quad \quad
\end{eqnarray}

In Table 6 we compare between the first-order approximate
solutions given by Eq. (\ref{AliM82}) with numerical results. The
corresponding relative errors are also presented.

\begin{table}[h!]
\caption{Comparison between OHAM results given by Eq.
(\ref{AliM82}) and numerical results for $f_w = 0$,
$\Lambda = 1$, $c = \frac{1}{2}$, $n= 1$, $Pr=2$ }%
\label{tab:6}
\begin{tabular}{lllll}
 \hline\noalign{\smallskip}
  $\eta$ & ${\theta}_{\textrm{numeric}}$ & $ \overline{\theta}_{\textrm{OHAM}}$, Eq. (\ref{AliM82})& $\begin{array}{c}
                                                         \textrm{relative error}= \\
                                                          |{\theta}_{\textrm{numeric}}-\overline{\theta}_{\textrm{OHAM}}|
                                                          \end{array}$\\
   \noalign{\smallskip}\hline\noalign{\smallskip}
0   &   1              &   1         &   0   \\   
1   &   0.1197281855   &   0.1187072441   &   1.02 $\cdot 10^{-3}$ \\   
2   &   0.0042249398   &   0.0041010249   &   1.23 $\cdot 10^{-4}$ \\   
3   &   5.09 $\cdot 10^{-5}$    &   -6.04 $\cdot 10^{-6}$     &   5.69 $\cdot 10^{-5}$ \\   
4   &   2.41 $\cdot 10^{-7}$   &    0.0001841967   &   1.83 $\cdot 10^{-4}$ \\   
5   &   1.67 $\cdot 10^{-8}$   &    0.0000163574   &   1.63 $\cdot 10^{-5}$ \\   
6   &   1.42 $\cdot 10^{-8}$   &   -0.0001452859   &   1.45 $\cdot 10^{-4}$ \\   
7   &   1.26 $\cdot 10^{-8}$   &   -0.0001791057   &   1.79 $\cdot 10^{-4}$ \\   
8   &   1.15 $\cdot 10^{-8}$   &   -0.0001403309   &   1.40 $\cdot 10^{-4}$ \\   
9   &   1.09 $\cdot 10^{-8}$   &   -0.0000887807   &   8.87 $\cdot 10^{-5}$ \\   
10  &   1.05 $\cdot 10^{-8}$   &   -0.0000493843   &   4.93 $\cdot 10^{-5}$ \\    %
\noalign{\smallskip}\hline
    \end{tabular}
\end{table}


{\textbf{Example 5.3.a}} We consider $f_w=1$, $\Lambda=1$,
$c=\frac{1}{2}$, $n=1$, $Pr=0.7$. The convergence-control
parameters for Eq. (\ref{AliM46}) are given by:
\begin{eqnarray}
   C_1 = 0.6287723857, \ C_2 =  -0.1379103919, \
C_3 =  -64.6127509553, \nonumber \\
 C_4 =  52.1862259014, \
 C_5 =  65.7049797786, \
 C_6 =  66.9471031457,  \nonumber \\
    K =  1.6976766716.  \nonumber
\end{eqnarray}
The first-order approximate solution (\ref{AliM46}) one can put
as:
\begin{eqnarray}\label{AliM83}
\overline{f}(\eta) = 1.6119245343 + (-0.1095284867  - 0.0145745495
\eta +  \nonumber \\
    0.0381089526 \eta^2
    -  0.0067695650 \eta^3) e^{-1.6976766716 \eta} +  (-0.6841717456
    +  \nonumber \\
    + 0.0590171180 \eta
-    1.5089146740 \eta^2) e^{-3.3953533432 \eta} +  (0.1817756979
-   \nonumber \\
- 0.6276022634 \eta -
    0.4839278208 \eta^2) e^{-5.0930300148 \eta}  \quad \quad
\end{eqnarray}

The convergence-control parameters for Eq. (\ref{AliM59}),  are:
\begin{eqnarray}
   C_7 = -2.4317156290, \
  C_8 =   0.4199773974, \
   C_9 =  3.2538989273,  \nonumber \\
C_{10} =  2.8819189890, \ C_{11} = 2.6307320394, \
  C_{12}= 0.0503403055  \nonumber
\end{eqnarray}
and the first-order approximate solution (\ref{AliM59}) becomes:
\begin{eqnarray}\label{AliM84}
\overline{\theta}(\eta) = (0.0191058146 +
    1.8994242629 \eta      - 0.7216780879 \eta^2 +  \nonumber \\
+    0.0699962329 \eta^3) e^{-1.6976766716 \eta} + (0.9928667494 +  \nonumber \\
+    0.3712879543 \eta -
    0.4243916166 \eta^2) e^{-3.3953533432 \eta} + (-0.0119725641 -  \nonumber \\
    - 0.1259716711 \eta -
    0.0024710391 \eta^2) e^{-5.0930300148 \eta}
    \quad \quad
\end{eqnarray}

In Tables 7 and 8 we present a comparison between the first-order
approximate solutions given by Eqs. (\ref{AliM83}) and
(\ref{AliM84}) respectively, with numerical results and
corresponding relative errors.

\begin{table}[h!]
\caption{Comparison between OHAM results given by Eq.
(\ref{AliM83}) and numerical results for $f_w = 1$,
$\Lambda = 1$ }%
\label{tab:7}
\begin{tabular}{lllll}
 \hline\noalign{\smallskip}
  $\eta$ & ${f}_{\textrm{numeric}}$ & $ \overline{f}_{\textrm{OHAM}}$, Eq. (\ref{AliM83})& $\begin{array}{c}
                                                         \textrm{relative error}= \\
                                                          |{f}_{\textrm{numeric}}-\overline{f}_{\textrm{OHAM}}|
                                                          \end{array}$\\
   \noalign{\smallskip}\hline\noalign{\smallskip}
0   &   1              &   1.00           &   2.22 $\cdot 10^{-16}$ \\   
1   &   1.5177074192   &   1.5176780223   &   2.93 $\cdot 10^{-5}$ \\   
2   &   1.6030516967   &   1.6030350430   &   1.66 $\cdot 10^{-5}$ \\   
3   &   1.6114161917   &   1.6114348208   &   1.86 $\cdot 10^{-5}$ \\   
4   &   1.6119056438   &   1.6119031833   &   2.46 $\cdot 10^{-6}$ \\   
5   &   1.6119228465   &   1.6119073012   &   1.55 $\cdot 10^{-5}$ \\   
6   &   1.6119232066   &   1.6119136285   &   9.57 $\cdot 10^{-6}$ \\   
7   &   1.6119232084   &   1.6119199331   &   3.27 $\cdot 10^{-6}$ \\   
8   &   1.6119232063   &   1.6119229505   &   2.55 $\cdot 10^{-7}$ \\   
9   &   1.6119232045   &   1.6119240509   &   8.46 $\cdot 10^{-7}$ \\   
10  &   1.6119232031   &   1.6119243982   &   1.19 $\cdot 10^{-6}$ \\   
\noalign{\smallskip}\hline
    \end{tabular}
\end{table}

\begin{table}[h!]
\caption{Comparison between OHAM results given by Eq.
(\ref{AliM84}) and numerical results for $f_w =1$,
$\Lambda = 1$, $c = \frac{1}{2}$, $n= 1$, $Pr=0.7$ }%
\label{tab:8}
\begin{tabular}{lllll}
 \hline\noalign{\smallskip}
  $\eta$ & ${\theta}_{\textrm{numeric}}$ & $ \overline{\theta}_{\textrm{OHAM}}$, Eq. (\ref{AliM84})& $\begin{array}{c}
                                                         \textrm{relative error}= \\
                                                          |{\theta}_{\textrm{numeric}}-\overline{\theta}_{\textrm{OHAM}}|
                                                          \end{array}$\\
   \noalign{\smallskip}\hline\noalign{\smallskip}
0   &   1              &   1              &   0 \\   
1   &   0.2625870984   &   0.2626175913   &   3.04 $\cdot 10^{-5}$ \\   
2   &   0.0500304293   &   0.0500306615   &   2.32 $\cdot 10^{-7}$ \\   
3   &   0.0067456425   &   0.0067634155   &   1.77 $\cdot 10^{-5}$ \\   
4   &   0.0006411532   &   0.0006125219   &   2.86 $\cdot 10^{-5}$ \\   
5   &   0.0000429270   &   0.0000457402   &   2.81 $\cdot 10^{-6}$ \\   
6   &   2.01 $\cdot 10^{-6}$    &    2.08 $\cdot 10^{-5}$   &  1.88 $\cdot 10^{-5}$ \\   
7   &   5.14 $\cdot 10^{-8}$    &    1.35 $\cdot 10^{-5}$   &  1.34 $\cdot 10^{-5}$ \\   
8   &   -1.29 $\cdot 10^{-8}$   &   6.14 $\cdot 10^{-6}$   &   6.16 $\cdot 10^{-6}$ \\   
9   &   -1.33 $\cdot 10^{-8}$   &   2.24 $\cdot 10^{-6}$   &   2.25 $\cdot 10^{-6}$ \\   
10  &   -1.22 $\cdot 10^{-8}$   &   7.13 $\cdot 10^{-7}$   &   7.25 $\cdot 10^{-7}$ \\   
\noalign{\smallskip}\hline
    \end{tabular}
\end{table}


{\textbf{Example 5.3.b}} For $f_w=1$, $\Lambda=1$,
$c=\frac{1}{2}$, $n=1$, $Pr=2$ the first-order approximate
solution for $\overline{f}(\eta)$ is given by Eq. (\ref{AliM83}).

For Eq. (\ref{AliM59}) the convergence-control parameters are
given by:
\begin{eqnarray}
      C_7 = -0.3023817542, \
      C_8 =  0.0519503347, \
     C_9 = -27.1653468182,  \nonumber \\
   C_{10} = 23.1898068679, \
 C_{11} =  -15.3543205956, \
   C_{12} = 31.0226487950. \nonumber
\end{eqnarray}
The first-order approximate solution (\ref{AliM59}) one retrieves
as:
\begin{eqnarray}\label{AliM85}
\overline{\theta}(\eta) = (0.9205703595 -  0.1921602046 \eta -
    0.0487183578 \eta^2 + \nonumber \\
 +  0.0086583891 \eta^3) e^{-1.6976766716 \eta} + (0.2637846962 +  \nonumber \\
+    0.9429053366 \eta -
    3.4149327807 \eta^2) e^{-3.3953533432 \eta}   + (-0.1843550558 -  \nonumber \\
    - 2.0986586159 \eta -
    1.5227992326 \eta^2) e^{-5.0930300148 \eta}  \quad \quad
\end{eqnarray}

In Table 9 we present a comparison between the first-order
approximate solutions given by Eqs. (\ref{AliM85}) with numerical
results. The corresponding relative errors are presented.

\begin{table}[h!]
\caption{Comparison between OHAM results given by Eq.
(\ref{AliM85}) and numerical results for $f_w = 1$,
$\Lambda = 1$, $c = \frac{1}{2}$, $n= 1$, $Pr=2$ }%
\label{tab:9}
\begin{tabular}{lllll}
 \hline\noalign{\smallskip}
  $\eta$ & ${\theta}_{\textrm{numeric}}$ & $ \overline{\theta}_{\textrm{OHAM}}$, Eq. (\ref{AliM85})& $\begin{array}{c}
                                                         \textrm{relative error}= \\
                                                          |{\theta}_{\textrm{numeric}}-\overline{\theta}_{\textrm{OHAM}}|
                                                          \end{array}$\\
   \noalign{\smallskip}\hline\noalign{\smallskip}
0   &   1              &   1.00           &    2.22 $\cdot 10^{-16}$ \\   
1   &   0.0288461240   &   0.0286378502   &    2.08 $\cdot 10^{-4}$ \\   
2   &   0.0002627867   &   0.0004342087   &    1.71 $\cdot 10^{-4}$ \\   
3   &   1.00 $\cdot 10^{-6}$  &    - 1.91 $\cdot 10^{-4}$  &    1.91 $\cdot 10^{-4}$ \\   
4   &   1.21 $\cdot 10^{-7}$   &   - 1.46 $\cdot 10^{-4}$  &    1.46 $\cdot 10^{-4}$ \\   
5   &   1.05 $\cdot 10^{-7}$  &   - 3.96 $\cdot 10^{-5}$  &    3.97 $\cdot 10^{-5}$ \\   
6   &   9.39 $\cdot 10^{-8}$   &   - 4.54 $\cdot 10^{-6}$   &   4.63 $\cdot 10^{-6}$ \\   
7   &   8.68 $\cdot 10^{-8}$   &   1.08 $\cdot 10^{-6}$    &   9.96 $\cdot 10^{-7}$ \\   
8   &   7.95 $\cdot 10^{-8}$    &   8.82 $\cdot 10^{-7}$   &   8.02 $\cdot 10^{-7}$ \\   
9   &   7.44 $\cdot 10^{-8}$   &   3.60 $\cdot 10^{-7}$   &   2.85 $\cdot 10^{-7}$ \\   
10  &   6.91 $\cdot 10^{-8}$   &   1.18 $\cdot 10^{-7}$   &   4.89 $\cdot 10^{-8}$  \\   
\noalign{\smallskip}\hline
    \end{tabular}
\end{table}
\begin{tabular}[!h]{c c}
  \includegraphics[width=2.2in]{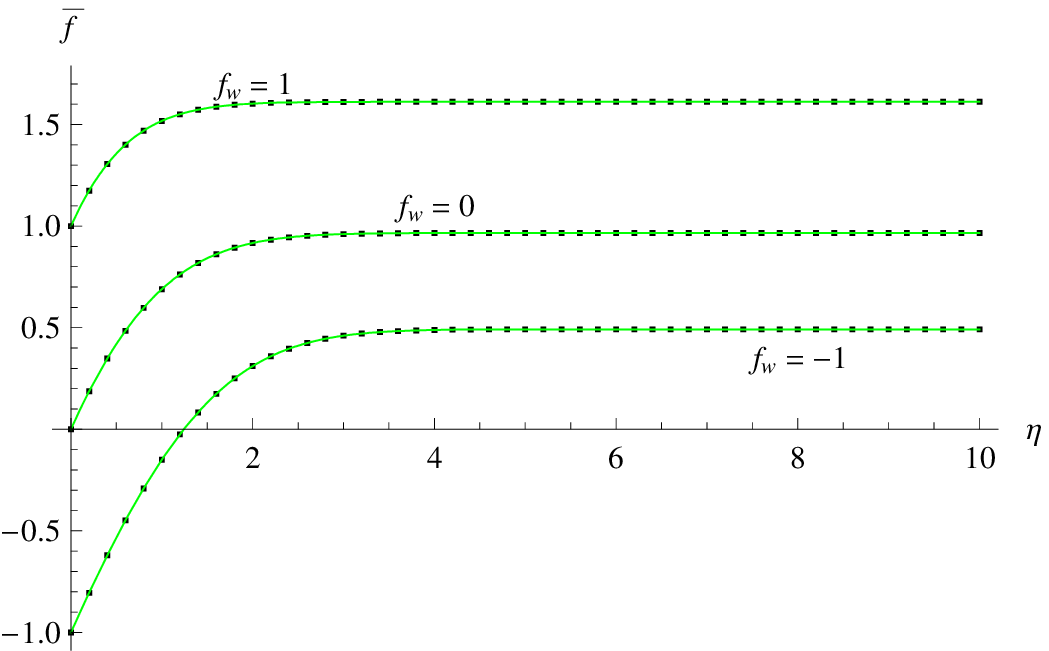}& \includegraphics[width=2.2in]{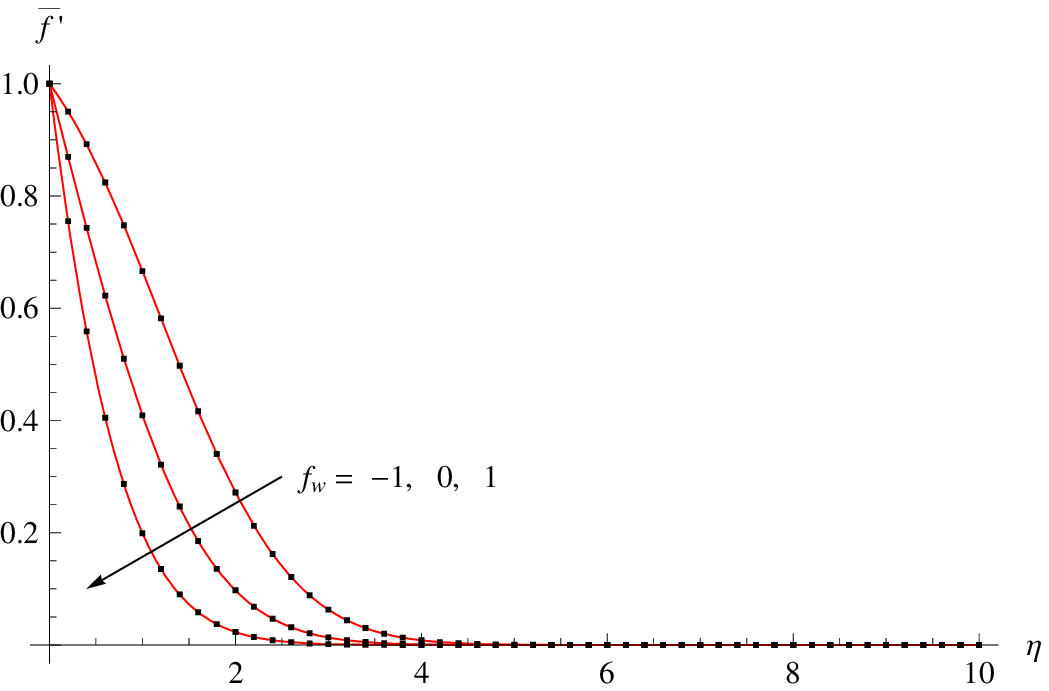} \\
  $\begin{array}{c}
  \textrm{Fig. 1 Solutions} \ \overline{f}_{OHAM}(\eta) \ \textrm{given by}\\
  \textrm{(\ref{AliM77}), (\ref{AliM80}) and (\ref{AliM83})} \ \\
 \textrm{for different values of} \ f_w\\
\textrm{--- numerical solution;} \\
\textrm{ ...... OHAM solution}
\end{array}$ & $\begin{array}{c}
 \textrm{Fig. 2 Solutions} \ {\overline{f}'}_{OHAM}(\eta) \ \\
  \textrm{ obtained from (\ref{AliM77}), (\ref{AliM80}) and (\ref{AliM83})} \ \\
 \textrm{for different values of} \ f_w\\
\textrm{--- numerical solution;} \\
\textrm{ ...... OHAM solution}
  \end{array}$ \\
\end{tabular}\\
\begin{tabular}[!t]{c c}
  \includegraphics[width=2.2in]{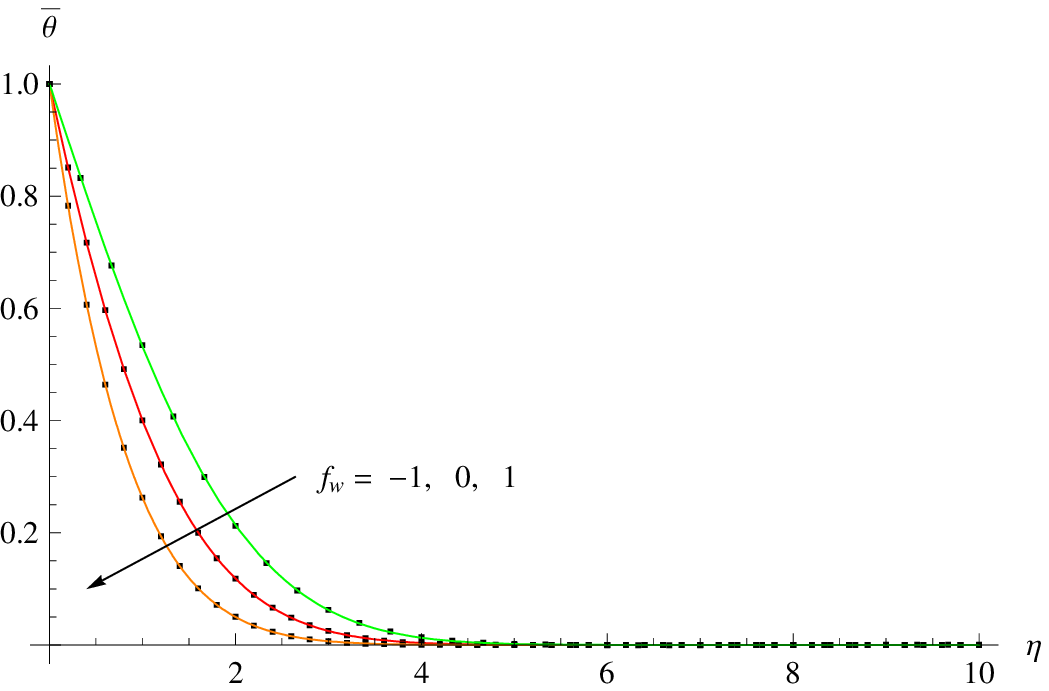}& \includegraphics[width=2.2in]{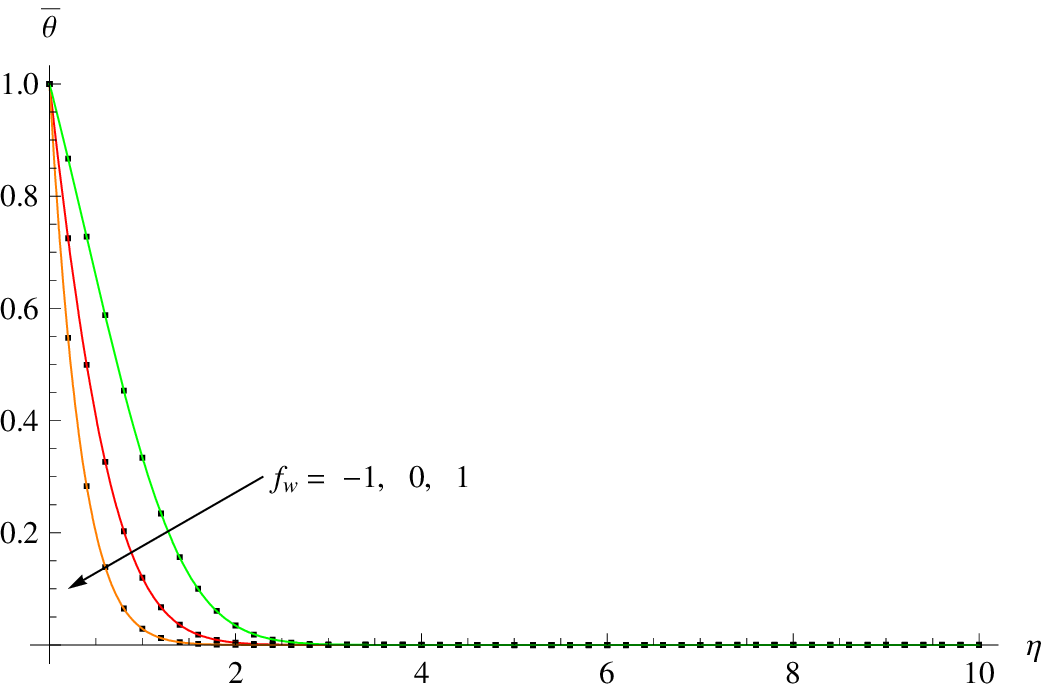} \\
  $\begin{array}{c}
  \textrm{Fig. 3 Plots of} \ \overline{\theta}_{OHAM}(\eta) \ \textrm{given}\\
  \textrm{ by Eqs. (\ref{AliM78}), (\ref{AliM81}) and (\ref{AliM84})} \ \\
 \textrm{for} \ \Lambda=1, \ c=\frac{1}{2}, \ n=1, \ Pr=0.7\\
\textrm{and three values of} \ f_w\\
\textrm{--- numerical solution;} \\
\textrm{ ...... OHAM solution}
\end{array}$ & $\begin{array}{c}
  \textrm{Fig. 4 Plots of} \ \overline{\theta}_{OHAM}(\eta) \ \textrm{given }\\
  \textrm{ by Eqs. (\ref{AliM79}), (\ref{AliM82}) and (\ref{AliM85})} \ \\
 \textrm{for} \ \Lambda=1, \ c=\frac{1}{2}, \ n=1, \ Pr=2\\
\textrm{and three values of} \ f_w\\
\textrm{--- numerical solution;} \\
\textrm{ ...... OHAM solution}
  \end{array}$ \\
\end{tabular}\\

In Figs 1 and 2 are plotted the profiles of $\overline{f}(\eta)$
and velocity profile $\overline{f}'(\eta)$ respectively for
different values of $f_w$. It is clear that the solution
$\overline{f}(\eta)$ increases with an increase of $f_w$ and the
velocity decrease with an increase of $f_w$. The condition
$\overline{f}'(\eta) > 0$ for $\eta>0$ is satisfied.\\

In Figs. 3 - 7 are plotted the temperature profiles given for two
values of Prandl number $Pr=0.7$ and $Pr=2$ respectively and
different values of $f_w$. From Figs 3 and 4 it is observe that
the temperature $\overline{\theta}(\eta)$ decreases with an
increase of the $f_w$ for any values of parameter $Pr$.\\

From Figs. 5-7 we can conclude that the temperature decrease with
of the Prandl number and different values of $f_w$.\\
\begin{tabular}[!t]{c c}
  \includegraphics[width=2.2in]{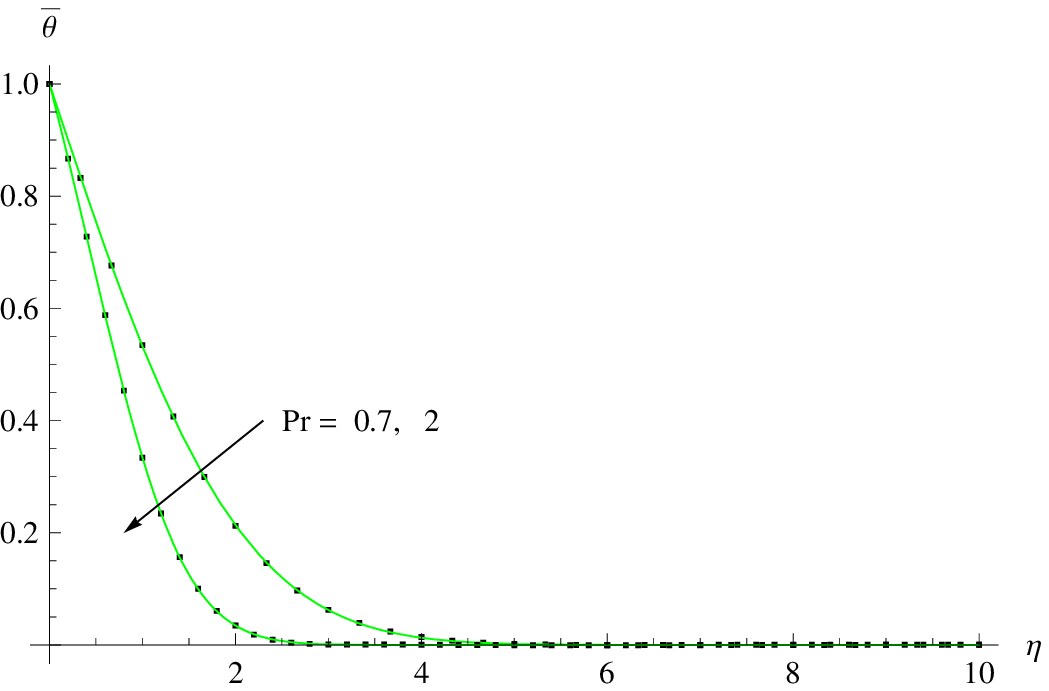}& \includegraphics[width=2.2in]{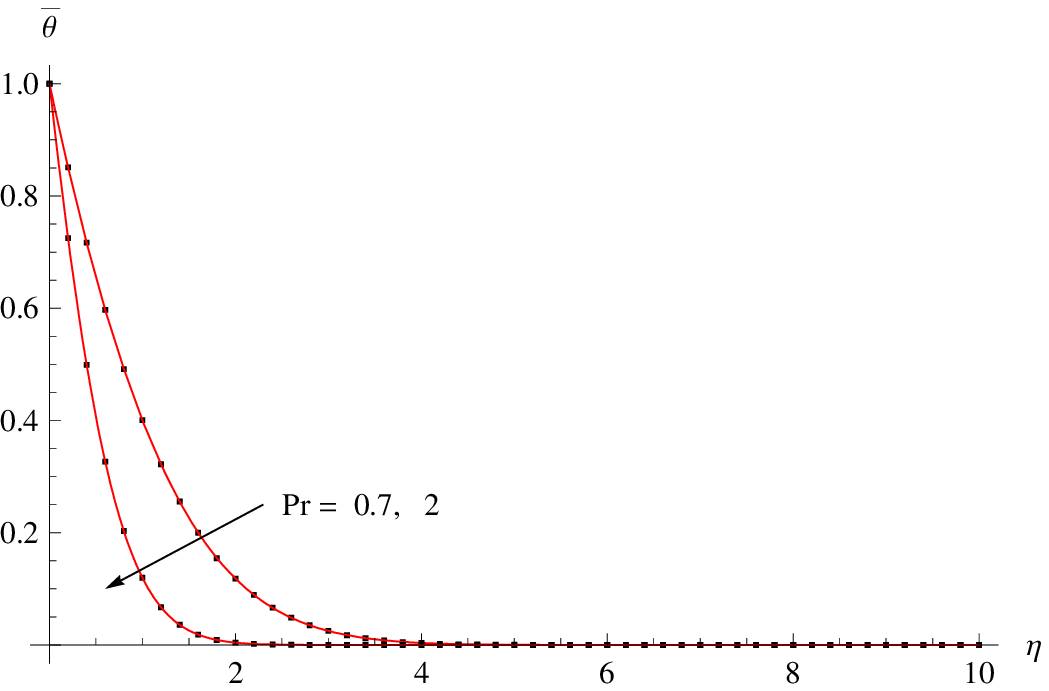} \\
  $\begin{array}{c}
  \textrm{Fig. 5 Plots of} \ \overline{\theta}_{OHAM}(\eta)  \\
  \textrm{given by Eqs. (\ref{AliM78}) and (\ref{AliM79})} \\
 \textrm{for} \ \Lambda=1, \ c=\frac{1}{2}, \ n=1, \\ f_w=-1 \
\textrm{and two values of} \ Pr\\
\textrm{--- numerical solution;} \\
\textrm{ ...... OHAM solution}
\end{array}$ & $\begin{array}{c}
  \textrm{Fig. 6 Plots of} \ \overline{\theta}_{OHAM}(\eta) \\
  \textrm{given by Eqs. (\ref{AliM81}) and (\ref{AliM82})} \\
 \textrm{for} \ \Lambda=1, \ c=\frac{1}{2}, \ n=1, \\ f_w=0 \
\textrm{and two values of} \ Pr\\
\textrm{--- numerical solution;} \\
\textrm{ ...... OHAM solution}
  \end{array}$ \\
\end{tabular}\\
\begin{tabular}[!t]{c}
  \includegraphics[width=2.2in]{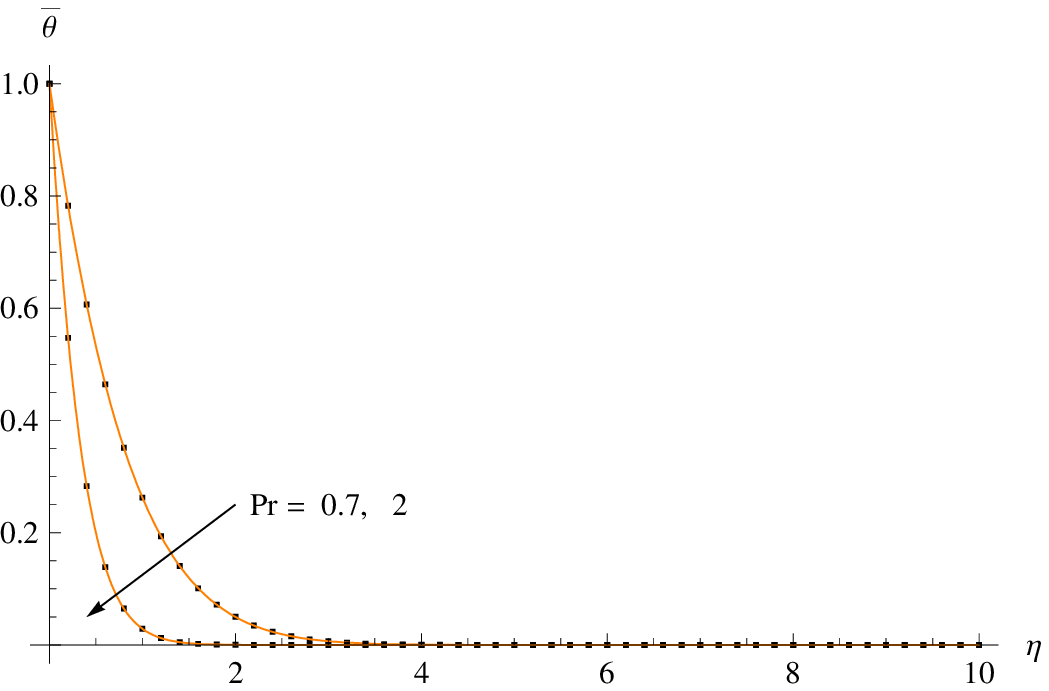} \\
  $\begin{array}{l}
  \textrm{Fig. 7 Plots of} \ \overline{\theta}_{OHAM}(\eta) \ \textrm{given by} \
  \textrm{Eqs. (\ref{AliM84}) and (\ref{AliM85})} \
 \textrm{for} \ \Lambda=1, \ c=\frac{1}{2}, \\
  n=1, \ f_w=1 \
\textrm{and two values of} \ Pr: \ \ \textrm{--- numerical
solution;} \\
\textrm{ ...... OHAM solution}
\end{array}$
\end{tabular}


From Tables 1-9 
we can summarize that the results obtained by means of OHAM are
very accurate in comparison with the numerical results.


\section{Conclusions}
\label{sec:6}

In this work, the Optimal Homotopy Asymptotic Method (OHAM) is
employed to propose analytical approximate solutions to the flow
and heat transfer in a viscous fluid over an unsteady stretching
surface. For three values of the suction/injection parameter
$f_w$, the problem admits solutions which are compared with
numerical solutions computed by means of the shooting method
combined with Runge-Kutta method and using Wolfram Mathematica 6.0
software. An analytical expressions for the heat transfer for two
values of the Prandl number are obtained. The solution
$\overline{f}(\eta)$ increases with an increase of $f_w$ and
velocity decrease with an increase of $f_w$. The temperature
$\overline{\theta}(\eta)$ decreases monotonically with the Prandl
number and with the distance $\eta$ from the stretching surface.

Our procedure is valid even if the nonlinear equations of the
motion do not contain any small or large parameters. The proposed
approach is mainly based on a new construction of the solutions
and especially on the involvement of the convergence-control
parameters via the auxiliary functions. These parameters lead to
an excellent agreement of the solutions with numerical results.
This technique is very effective, explicit and accurate for
nonlinear approximations rapidly converging to the exact solution
after only one iteration. Also, OHAM provides a simple but
rigorous way to control and adjust the convergence of the solution
by means of some convergence-control parameters. Our construction
of homotopy is different from other approaches especially
referring to the linear operator $L$ and to the auxiliary
convergent-control function $H_f$ and $H_{\theta}$ which ensure a
fast convergence of the solutions.

It is worth mentioning that the proposed method is
straightforward, concise and can be applied to other nonlinear
problems.

\section*{Acknowledgement} The authors declare that there is no
conflict of interests regarding the publication of this paper.

\end{document}